\input harvmac.tex
\vskip 1.5in
\Title{\vbox{\baselineskip12pt
\hbox to \hsize{\hfill}
\hbox to \hsize{\hfill}}}
{\vbox{\centerline{New Discrete States
in Two-Dimensional Supergravity}
\vskip 0.3in
{\vbox{\centerline{}}}}}
\centerline{Dimitri Polyakov\footnote{$^\dagger$}
{dp02@aub.edu.lb}}
\medskip
\centerline{\it Center for Advanced Mathematical Sciences}
\centerline{\it and  Department of Physics }
\centerline{\it  American University of Beirut}
\centerline{\it Beirut, Lebanon}
\vskip .2in
\centerline {\bf Abstract}
Two-dimensional string theory is known to contain the set of
discrete states that are the
SU(2) multiplets generated by the lowering operator
of the $SU(2)$ current algebra.
Their structure constants are defined by the area preserving diffeomorphisms
in two dimensions.
In this paper we show that the 
interaction of $d=2$ superstrings
with the superconformal $\beta-\gamma$
ghosts enlarges the actual
 algebra of the dimension 1 currents and hence
the new ghost-dependent discrete states 
appear. Generally, these states are the SU(N) multiplets, if the algebra
includes the currents of ghost numbers n : $-N\leq{n}\leq{N-2}$,
not related by picture-changing. We compute the structure constants
of these ghost-dependent discrete states for $N=3$ and
express them in terms of $SU(3)$ Clebsch-Gordan coefficients,
relating this operator algebra
to the volume preserving diffeomorphisms
in $d=3$. For general $N$, the operator algebra is conjectured to be 
isomorphic to $SDiff(N)$. This points at possible holographic
relations between two-dimensional superstrings and field theories 
in higher dimensions.
 {\bf}
\Date{February 2006}
\vfill\eject 
\lref\lian{B. Lian, G. Zuckerman, Phys.Lett. B254 (1991) 417}
\lref\pol{I. Klebanov, A. M. Polyakov, Mod.Phys.Lett.A6:3273-3281}
\lref\wit{E. Witten, Nucl.Phys.B373:187-213  (1992)}
\lref\grig{M. Grigorescu, math-ph/0007033, Stud. Cercetari Fiz.36:3 (1984)}
\lref\witten{E. Witten,hep-th/0312171, Commun. Math. Phys.252:189  (2004)}
\lref\ohta{K.Itoh, N.Ohta,Nucl.Phys. B377 (1992) 113}
\lref\itoh{K.Itoh, N.Ohta, Prog. Th. Phys.Suppl. 110 (1992) 97} 
\lref\wb{N. Berkovits, E. Witten, hep-th/0406051, JHEP 0408:009 (2004)}
\lref\zam{A. Zamolodchikov and Al. Zamolodchikov,
Nucl.Phys.B477, 577 (1996)}
\lref\mars{J. Marsden, A. Weinstein, Physica 7D (1983) 305-323}
\lref\arnold{V. I. Arnold,''Geometrie Differentielle des Groupes de Lie'',
Ann. Inst. Fourier, Grenoble 16, 1 (1966),319-361}
\lref\self{D. Polyakov,  Int. Jour. Mod. Phys A20: 2603-2624 (2005)}
\lref\selff{D. Polyakov, Phys. Rev. D65: 084041 (2002)} 
\lref\sellf{D. Polyakov, Int. J. Mod. Phys A20:4001-4020 (2005)}
\lref\doug{M.Douglas et.al. , hep-th/0307195}
\lref\dorn{H. Dorn, H. J. Otto, Nucl. Phys. B429,375 (1994)}
\lref\prakash{J. S. Prakash, 
H. S. Sharatchandra, J. Math. Phys.37:6530-6569 (1996)}
\centerline{\bf Introduction}
The spectrum of physical states of non-critical
one-dimensional string theory (or , equivalently, of critical
string theory in two dimensions) is known to contain the
``discrete states'' of non-standard b-c ghost numbers 0 and 2
(while the standard unintegrated vertex operators 
always carry the ghost number 1) ~{\lian, \pol,\wit,\ohta,\itoh} 
These states appear at
the special (integer or half integer) values of the momentum $p$ but
are absent for generic $p$ {\footnote{$^1$} {in our paper
 the stress tensor for the $X$-field 
is normalized as $T=-{1\over2}(\partial{X})^2$
which is one half of the normalization used in ~{\pol}
hence only the integer values of p are allowed}}

The appearance of
these states is closely related to the  $SU(2)$ symmetry,
 emerging when the theory is compactified on a circle with
the self-dual radius ~{\pol, \wit}

The SU(2) symmetry is generated by 3 currents of 
conformal dimension 1:
\eqn\grav{\eqalign{T_3=\oint{{dz}\over{2i\pi}}\partial{X}(z)\cr
T_{+}=\oint{{dz}\over{2i\pi}}e^{{iX}{\sqrt{2}}}\cr
T_{-}=\oint{{dz}\over{2i\pi}}e^{{-iX}{\sqrt{2}}}}}
 The tachyonic operators $W_{s}=
e^{{isX}\over{\sqrt{2}}}$
with  non-negative integer $s$ are the
highest weight vestors in the representation of the ``angular momentum'' $s$ 
 (here X is the $c=1$ matter field).

The discrete primaries are then constructed
by repeatedly acting on $W_s$ with the lowering operator
$T_{-}$ of SU(2). For example,  acting on $W_{s}$ n times
with $T_{-}$
one generally produces the primary fields
of the form
$W_{s,n}=P(\partial{X},\partial^{2}X,...)
e^{i({{s}\over{\sqrt{2}}}-n\sqrt{2})X}$ with
$n<s$,
where $P$ is some polynomial in the derivatives of $X$, straightforward but
generally hard to compute.
After the Liouville dressing and the multiplication
 by the ghost field $c$, 
$W_{s,n}$ becomes the dimension 0 primary field
$cW_{s,n}$, i.e. the BRST-invariant operator for a physical state.
The set of operators $cW_{s,n}$
 thus consists of the multiplet states of $SU(2)$
which OPE structure constants  can be shown to form
the enveloping of $SU(2)$ ~{\pol}.
The discrete operators of non-standard adjacent ghost numbers 0 and 2
$Y_{s,n}^{0}$ and $Y_{s,n}^{+2}$
 can
be obtained from $cW_{s,n}$ simply by considering the  BRST commutators
~{\wit}:
$\lbrace{Q_{BRST},Y_{s,n}^{0}\rbrace}=cW_{s,n}$ and
$\lbrace{Q_{BRST},cW_{s,n}}\rbrace=Y^{+2}_{s,n}$.
One has to
evaluate these commutators
 for arbitrary $s$ first and and then take $s$ to be
the appropriate  integer number.
For arbitrary $s$ the ghost number 0 and 2 
Y-operators are respectively BRST non-invariant and exact,
but for the integer values of $s$ they become independent
physical states, defining the BRST cohomologies with non-trivial $b-c$
 ghost numbers.
At the first glance, things naively could seem to be the same in case when
the $c=1$ theory is supersymmetrized on the worldsheet.
However, in this paper we will show that the interaction of the
$c=1$ theory with the system of $\beta-\gamma$ ghosts dramatically
extends the spectrum of the physical discrete states.
Consider the $c=1$ model supersymmetrized
on the worldsheet coupled to the super Liouville field.
The worldsheet action of the system in the conformal gauge is given by
~{\doug}
\eqn\grav{\eqalign{
S=S_{X-\psi}+S_{L}+S_{b-c}+S_{\beta-\gamma}\cr
S_{X-\psi}={1\over{4\pi}}\int{d^2z}
\lbrace\partial{X}\bar\partial{X}+\psi\bar\partial\psi
+\bar\psi\partial\bar\psi\rbrace\cr
S_{L}={1\over{4\pi}}\int{d^2z}\lbrace\partial\varphi\bar\partial\varphi
+\lambda\bar\partial\lambda+\bar\lambda\partial\bar\lambda
-F^2+2\mu_0be^{b\varphi}(ib\lambda\bar\lambda-F)\rbrace\cr
S_{b-c}+S_{\beta-\gamma}={1\over{4\pi}}
\int{d^2z}\lbrace{b}\bar\partial{c}+{\bar{b}}\partial{\bar{c}}
+\beta\bar\partial\gamma+\bar\beta\partial\bar\gamma\rbrace}}
with $Q\equiv{b+b^{-1}}$ being the  background charge.

The stress tensors of the matter and the ghost systems
and the standard bosonization relations for the ghosts are given by
\eqn\grav{\eqalign{T_{m}=-{1\over2}(\partial{X})^2
-{1\over2}\partial\psi\psi-{1\over2}(\partial\varphi)^2
+{{Q}\over2}\partial^2\varphi\cr
T_{gh}={1\over2}(\partial{\sigma})^2+{3\over2}\partial^2\sigma
+{1\over2}(\partial\chi)^2+{1\over2}\partial^2\chi
-{1\over2}(\partial\phi)^2-\partial^2\phi\cr
c=e^\sigma,b=e^{-\sigma}\cr
\gamma=e^{\phi-\chi},\beta=e^{\chi-\phi}\partial\chi}}
The central charge of the super Liouville system
is equal to $c_L=1+2Q^2$ and, since
$c_{X-\psi}+c_L+c_{b-c}+c_{\beta-\gamma}=0$,
we have $Q^2=4$ for one-dimensional non-critical
superstrings.

The $SU(2)$ algebra is now generated by
the dimension 1 currents
\eqn\grav{\eqalign{T_{0,0}=\oint{{dz}\over{2i\pi}}
\partial{X}\cr
T_{0,1}=\oint{{dz}\over{2i\pi}}
e^{iX}\psi\cr
T_{0,-1}=\oint{{dz}\over{2i\pi}}e^{-iX}\psi}}
Here and elsewhere 
the first lower index refers to the superconformal ghost number
and the second to the momentum in the $X$ direction.
In comparison with the bosonic case,
a crucial novelty emerges  due to the interaction of the matter
with the $\beta-\gamma$ system of superconformal ghosts.
That is, we will show that, apart from the
dimension one currents (4) generating SU(2),there is also a set of
 BRST-nontrivial dimension 1 currents mixed with $\beta-\gamma$ ghosts,
i.e. those existing at nonzero $\beta-\gamma$ ghost pictures and
not  reducible to (4) by any picture-changing transformation.
These extra currents turn out to  enhance  the actual
underlying symmetry of the theory from $SU(2)$ two
$SU(n)$ with $n\geq{3}$ being the highest order ghost number
cohomology of the operators in the current algebra 
(see the discussion below).
For example, the currents of the form
\eqn\lowen{
T_{-n,n-1}=\oint{{dz}\over{2i\pi}}e^{-n\phi+(n-1)X}\psi(z)}
 with ghost numbers
$n\leq{-3}$ are BRST-nontrivial, even though they are annihilated by
picture-changing transformation and no picture zero version of these
currents exists. The possibility of existence of BRST-nontrivial operators
annihilated by picture-changing has been discussed in ~{\selff} in 
case of $n=-3$ for the critical $D=10$ superstrings.
In this paper we shall also present a proof of the BRST non-triviality
of  the  dimension 1 currents with arbitrary 
negative ghost numbers $n\leq{-3}$ 
in two-dimensional critical superstring theory.
Next, acting on $T_{-n,n-1}$ repeatedly  with $T_{-}$ of (4)
one obtains the currents of the form
\eqn\grav{\eqalign{T_{-n,m}=\oint{{dz}\over{2i\pi}}
P_{-n,m}(\partial{X},\partial^2{X},...;\psi,\partial\psi,...)
e^{-n\phi+imX}\cr
|m|\leq{n-1}}}
which generally are the BRST non-trivial Virasoro 
primaries of ghost number $-n$ annihilated by picture-changing.
Here $P_{-n,m}$ are the polynomials in $\partial{X},\psi$
and their derivatives, with the conformal weight
$h={1\over2}(n^2-m^2)+n+1$ so that the overall dimension
of the integrands in (6) is equal to 1.
In this paper we will derive the precise expressions
for these polynomials for the cases $n=-3$ and $n=-4$.
So if one considers all the picture-inequivalent currents with ghost numbers
$p=0,-3,....-n$, including those of (4),
 (all the dimension 1 currents with ghost numbers
-1 and -2  are equivalent to the $SU(2)$ generators (4) by picture-changing)
one has the total number $n^2-1$ of T-generators (6).
In this paper we will show (precisely for $|n|\leq{4}$ and conjecture
for larger values of $|n|$) that the algebra of operators $T_{p,m}$
$|m|\leq|p-1|,p=-3,...,-n$ combined with 3 $SU(2)$ generators of (4) 
is simply $SU(n)$ with  the generators
$T_{0,m}$ ($m=0;-3,-4,...-n$) being in the Cartan subalgebra of $SU(n)$.
The next step is easy to guess.
For each given n one starts with the Liouville
 dressed tachyonic Virasoro primaries
$$\oint{{dz}\over{2i\pi}}V_l=\oint{{dz}\over{2i\pi}}
e^{ilX+(l-1)\varphi}(l\psi-i(l-1)\varphi)$$, 
with integer $l$ and acts on them with
various combinations of lowering $T$-operators (i.e. those having singular
OPEs with integrands of $V_l$). The obtained operators will
be the multiplets of $SU(n)$, including the operators of BRST
cohomologies with non-trivial ghost dependence 
(not removable by the picture changing).
 These operators are generally of the form
\eqn\grav{\eqalign{V_{q,n;m,l}=\oint{{dz}\over{2i\pi}}
:{e^{(l-1)\varphi-q\phi+imX}}P_{q,n;m,l}
(\partial{X},\partial^2{X},...\psi,
\partial\psi,...,\partial\phi,\partial^2\phi...):(z)\cr
|m|\leq{l}}}
Of course,the usual well-known discrete states 
form the operator subalgebra of ghost cohomology number zero
 in the space of $\lbrace{V_{q,n;m,l}}\rbrace$.
Here and elsewhere the term ``ghost cohomology'' refers to
the factorization over all the picture-equivalent states;
thus the ghost cohomology of number $-q$ $(q>{0})$
consists of  BRST invariant and non-trivial operators
with the ghost numbers $-r\leq{-q}$ existing at the
maximal picture $-q$, so that their picture $-q$ expressions
are annihilated by the operation  of the picture-changing;
conversely, the operators of a positive ghost cohomology 
number $q>0$ are the physical operators annihilated by the 
inverse picture changing at the picture $q$,
so that they don't exist at pictures below $q$.
Of course, the ghost cohomologies are not  cohomologies
in the literal sense, since the picture-changing operator
is not nilpotent. 
In the rest of the paper we will particularly demonstrate 
the above construction by precise computations, deriving expressions for
the new vertex operators of the ghost-dependent discrete states and 
computing their structure constants.

\centerline{\bf Extended current algebra and ghost-dependent discrete states}

As usual, in the supersymmetric case 
three SU(2) currents (4) can be taken at different
ghost pictures. For example, the picture $-1$ expressions for the currents (4)
: $e^{-\phi}\psi,e^{-\phi\pm{iX}}$ are the only dimension 1 generators
at this superconformal ghost number; similarly, all the 
ghost number $-2$ dimension $1$ operators:
$e^{-2\phi}\partial{X},e^{-2\phi\pm{iX}}\psi$ are just the
the $SU(2)$ generators (4) at picture $-2$.
However, at  ghost pictures of $-3$ and below, the new
dimension $1$ generators,
not reducible to (4) by picture-changing, appear.
The first example is the generator given by the 
worldsheet integral 
\eqn\grav{\eqalign{T_{-3,2}=\oint{{dz}\over{2i\pi}}e^{-3\phi+2iX}\psi(z)}}
where as usual $\phi$ is a bosonized superconformal
ghost with the stress tensor 
$T_\phi=-{1\over2}(\partial\phi)^2-\partial^2\phi$.
This operator is annihilated by  the picture-changing transformation,
and we leave the proof of its
BRST non-triviality until the next section.
Given the generator (8), it is now straightforward to construct 
the dimension 1 currents
 of the ghost number $-3$ with the momenta $0,\pm{1}$ and $-2$,
which are the Virasoro primaries,not related to (4) by picture-changing.
For instance this can be done by taking the lowering operator
$T_{0,-1}=\oint{{dz}\over{2i\pi}}e^{-iX}\psi(z)$ of $SU(2)$
 and acting on (8).
Performing this simple calculation, we obtain  the 
following extra five generators in the ghost number $-3$ cohomology:
\eqn\grav{\eqalign{T_{-3,2}=\oint{{dz}\over{2i\pi}}e^{-3\phi+2iX}\psi(z)\cr
T_{-3,1}=\oint{{dz}\over{2i\pi}}e^{-3\phi+iX}(\partial\psi\psi
+{1\over2}(\partial{X})^2+{i\over2}\partial^2{X})(z)\cr
T_{-3,-1}=\oint{{dz}\over{2i\pi}}e^{-3\phi-iX}(\partial\psi\psi
+{1\over2}(\partial{X})^2-{i\over2}\partial^2{X})(z)\cr
T_{-3,0}=\oint{{dz}\over{2i\pi}}e^{-3\phi}(\partial^2{X}\psi-2\partial{X}
\partial{\psi})(z)\cr
T_{-3,-2}=\oint{{dz}\over{2i\pi}}e^{-3\phi-2iX}\psi(z)
}}
It is straightforward to check that all these generators
are the primary fields commuting with the BRST charge. 
The next step is to show that the operators (9) taken with  3
standard SU(2) generators $T_{0,0},T_{0,1}$
and $T_{0,-1}$ of (4)
 combine into  8 generators of $SU(3)$
(up to picture-changing transformations),
with $T_{0,0}$ and $T_{-3,0}$ generating the Cartan subalgebra
of $SU(3)$.
Here an important remark should be made.
Straightforward computation of  the commutators of
some of the generators (9) can
be quite cumbersome due to high order singularities in the OPE's
involving the exponential operators with negative ghost numbers
as $e^{\alpha\phi}(z)e^{\beta\phi}(w)\sim(z-w)^{-\alpha\beta}
e^{-(\alpha+\beta)\phi}(w)+...$
For instance, the computation of the commutator of $T_{-3,2}$ 
with $T_{-3,-2}$ would involve the OPE $e^{-3\phi+2iX}(z)e^{-3\phi-2iX}(w)
\sim(z-w)^{-13}e^{-6\phi}+...$, so the OPE coefficient in fromt of the
single pole would be cumbersome to compute. In addition,
as the ghost picture of the r.h.s. of the commutator is equal to $-6$
 one would need an additional picture-changing transformation
to relate it to the $T$-operators (4),(9). Things, however, 
can be simplified if we note that the operators $T_{-n,m}$
of ghost cohomology $-n$ ($n=3,4,...$) are picture-equivalent to the
appropriate
operators from the $positive$ ghost number $n-2$ cohomologies.  
That is, if in the expressions for any of these operators
one replaces $e^{-n\phi}$ with $e^{(n-2)\phi}$
(which has the same conformal dimension as $e^{-n\phi}$) and
keeps the matter part unchanged, one gets an equivalent 
BRST-invariant and nontrivial operator,
in the sense that replacing one operator with another inside any
correlator does not change the amplitude.
Such an equivalence is generally up to
certain b-c ghost terms needed to preserve the BRST-invariance of the 
operators with positive  superconformal ghost numbers; however, these terms, 
generally do not contribute to correlators due to the $b-c$ 
ghost number conservation conditions ~{\sellf}. 
Thus, for example, one can replace 
\eqn\lowen{T_{-3,2}=\oint{{dz}\over{2i\pi}}
{e^{-3\phi+2iX}}\psi\rightarrow
{T_{+1,2}}=\oint{{dz}\over{2i\pi}}e^{\phi+2iX}\psi} 
(up to the b-c ghost terms).
The picture-equivalence of these operators means that, even though 
they are not  straightforwardly related by picture-changing transformations
(as $T_{-3,2}$ and $T_{+1,2}$ are respectively annihilated by direct and
 inverse picture-changings), any correlation functions involving these 
vertex operators  are equivalent under the replacement (10).
The precise relation between these operators is given by
\eqn\lowen{T_{+1,2}=Z(::\Gamma^4:ce^{-3\phi+2iX}:)}
where $:\Gamma:^4$ is the normally ordered fourth power of the usual 
picture-changing operator $\Gamma$ while the $Z$-transformation mapping
the local vertices to integrated is the analogue of $\Gamma$ for the $b-c$
ghosts ~{\sellf}. The Z-transformation can be performed by
using the BRST-invariant non-local Z-operator of $b-c$ ghost
number $-1$;  the precise expression for this operator
was derived in ~{\sellf} and is given by
$$Z={\oint_{w}}{{du}\over{2i\pi}}(u-w)^3(bT+4ce^{2\chi-2\phi}T^2)(u)$$
where $T$ is the full matter $+$ ghost stress-energy tensor;
 the integral is taken over the worldsheet boundary and acts
on local operators at $w$.
Similarly, for any ghost number $-n$ generators one can replace
\eqn\lowen{T_{-n,m}\rightarrow{T_{n-2,m}};n\leq{-3}}
Since the Liouville and the $\beta-\gamma$ stress tensors
both have the same $Q^2$,
the equivalence relations (10), (12) can formally be thought of as the 
special case of the reflection identities  ~{\dorn,\zam}
in $d=2$ super Liouville theory 
 in the limit of zero cosmological constant $\mu_0$.
Using the equivalence identities (10),(12) and evaluating
the OPE simple poles in the commutators it is now
 straightforward to  determine the algebra of operators (4),(9).
At this point, it is convenient to redefine:
\eqn\grav{\eqalign{
L={i\over{2}}T_{0,0},H={i\over{3{\sqrt{2}}}}T_{-3,0}\cr
G_{+}={1\over{2\sqrt{2}}}({\sqrt{2}}T_{0,1}+T_{-3,1}),G_{-}
={1\over{2\sqrt{2}}}({\sqrt{2}}T_{0,1}-T_{-3,1})\cr
F_{+}={1\over{2{\sqrt{2}}}}({\sqrt{2}}T_{0,-1}+T_{-3,-1}),
F_{-}={1\over{2{\sqrt{2}}}}({\sqrt{2}}T_{0,-1}-T_{-3,-1})\cr
G_3={1\over{\sqrt{2}}}T_{-3,2},F_3={1\over{\sqrt2}}T_{-3,-2}
}}
Then the commutators of the operators $L$ and $H$ with each other and 
with the rest of (13) are  given by
\eqn\grav{\eqalign{
\lbrack{L,H}\rbrack=0\cr
\lbrack{L,G_{+}}\rbrack={1\over2}G_{+};
\lbrack{L,G_{-}}\rbrack={1\over2}G_{-};
\lbrack{L,F_{+}}\rbrack=-{1\over2}F_{+};
\lbrack{L,F_{-}}\rbrack=-{1\over2}F_{-}\cr
\lbrack{L,G_{3}}\rbrack=G_{3};
\lbrack{L,F_{3}}\rbrack=-F_{3}\cr
\lbrack{H,G_{+}}\rbrack=-G_{+};
\lbrack{H,G_{-}}\rbrack=G_{-};
\lbrack{H,F_{+}}\rbrack=F_{+};
\lbrack{H,F_{-}}\rbrack=-F_{-}\cr
\lbrack{H,G_{3}}\rbrack=\lbrack{H,F_3}\rbrack=0}}
i.e. $L$ and $H$ indeed are in the Cartan subalgebra;
the remaining commutators of the currents(13) 
are then easily computed to give
\eqn\grav{\eqalign{
\lbrack{G_3,G_{+}}\rbrack=0;
\lbrack{G_3,G_{-}}\rbrack=0;
\lbrack{G_3,F_{+}}\rbrack=G_{-};\lbrack{G_3,F_{-}}\rbrack=-{G}_{+}\cr
\lbrack{F_3,F_{+}}\rbrack=0;
\lbrack{F_3,F_{-}}\rbrack=0;
\lbrack{F_3,G_{+}}\rbrack=F_{-};\lbrack{F_3,G_{-}}\rbrack=-F_{+}\cr
\lbrack{F_{-},F_{+}}\rbrack=-F_3;
\lbrack{G_{-},G_{+}}\rbrack=-G_3;\lbrack{G_{-},F_{+}}\rbrack=
\lbrack{F_{-},G_{+}}\rbrack=0\cr
\lbrack{F_{+},G_{+}}\rbrack=L-{3\over2}H;\lbrack{F_{-},G_{-}}\rbrack=
L+{3\over2}H;\lbrack{G_{3},F_{3}}\rbrack=2L}}
Thus the operators $L,H,F_{\pm},G_{\pm},F_3$ and $G_3$ of (13)
simply define the Cartan-Weyl basis of $SU(3)$.
Their BRST invariance ensures that this $SU(3)$ algebra
intertwines with the superconformal symmetry of the theory.
Therefore, just like in the case of usual $SU(2)$ discrete states
of two-dimensional supergravity, we can generate the
extended set of discrete super Virasoro primaries by taking
a dressed tachyonic exponential operator $$V={\oint}{{dz}\over{2i\pi}}
e^{ilX+(l-1)\varphi}(l\psi-i(l-1)\lambda)$$
 at  integer values of $l$ and acting with
the  generators of the lowering subalgebra of
$SU(3)$ ($F_{\pm}$ and $F_3$).
The obtained physical operators are then the multiplets of
SU(3) and include
the new discrete physical states of non-trivial ghost cohomologies,
not reducible to the usual $SU(2)$ primaries
by any picture-changing transformations. 
Such a construction will be  demonstrated explicitly
 in the next section.
The scheme explained above can be easily generalized to
include the currents of higher values of ghost numbers.
For example, to include the currents of ghost numbers
up to $-4$ (or up to $+2$, given the equivalence relations(12))
one has to start with the
generator $T_{-4,3}=\oint{{dz}\over{2i\pi}}e^{-4\phi+3iX}\psi$.
The conformal dimension of the integrand is equal to 1.
This 
generator is not related to any of the operators
(4),(9)
 by the picture-changing and therefore is the part of the ghost 
 number $-4$ cohomology.
This cohomology contains 7 new currents $T_{-4,m},|m|\leq{3}$,
which are the BRST-invariant super Virasoro primaries.
 As previously, these currents can be  
generated from $T_{-4,3}$ by acting on it repeatedly with
$T_{0,-1}$ of (4).The resulting operators are given by
\eqn\grav{\eqalign{T_{-4,{\pm}3}=\oint{{dz}\over{2i\pi}}
e^{-4\phi{\pm}3iX}\psi(z)\cr
T_{-4,2}=\oint{{dz}\over{2i\pi}}e^{-4\phi+2iX}
({1\over2}\partial^2\psi\psi-{i\over6}\partial^3X+{i\over6}(\partial{X})^3
-{1\over2}\partial{X}\partial^2{X})\cr
T_{-4,1}=\oint{{dz}\over{2i\pi}}e^{-4\phi+iX}
({1\over2}\psi\partial{\psi}\partial^2\psi+{1\over{24}}
P_{-iX}^{(4)}\psi-{1\over4}P^{(2)}_{-iX}\partial^2\psi
-{1\over4}(P_{-iX}^{(2)})^2\psi)\cr
T_{-4,-1}=\oint{{dz}\over{2i\pi}}e^{-4\phi-iX}
({1\over2}\psi\partial{\psi}\partial^2\psi+{1\over{24}}
P_{iX}^{(4)}\psi-{1\over4}P^{(2)}_{iX}\partial^2\psi
-{1\over4}(P_{iX}^{(2)})^2\psi)\cr
T_{-4,-2}=\oint{{dz}\over{2i\pi}}e^{-4\phi-2iX}
({1\over2}\partial^2\psi\psi+{i\over6}\partial^3X-{i\over6}(\partial{X})^3
-{1\over2}\partial{X}\partial^2{X})\cr
T_{-4,0}=\oint{{dz}\over{2i\pi}}e^{-4\phi}
\lbrace{2i}\partial{X}\partial\psi\partial^2\psi
+P^{(2)}_{-iX}\psi\partial^2\psi-{2\over3}
P^{(3)}_{-iX}\psi\partial\psi\cr
-{1\over6}P^{(3)}_{-iX}P^{(2)}_{-iX}
+(\partial{X})^2\psi\partial^2\psi+{{7i}\over8}\partial{X}P^{(4)}_{-iX}\cr
-i(\partial{X})^3\psi\partial\psi-{i\over2}\partial{X}P^{(2)}_{-iX}
\psi\partial\psi+{i\over4}\partial{X}(P^{(2)}_{-iX})^2
-{1\over4}(\partial{X})^2P^{(3)}_{-iX}\rbrace}}

Here $P_{{\pm}iX}^{(n)};n=2,3,4$ are the 
conformal weight $n$ polynomials in the derivatives of $X$ defined as
\eqn\lowen{P_{f(X(z))}^{(n)}
=e^{-f(X(z))}{{\partial^n}\over{\partial{z}^n}}e^{f(X(z))}}
for any given function $f(X)$. 
For example, taking $f=iX$ one has
$P^{(1)}_{iX}=i\partial{X},P^{(2)}_{iX}=i\partial^2X-(\partial{X})^2$
etc.
The special case $f(X(z))=X(z)={{z^2}\over2}$ gives the usual Hermite
polynomials in $z$. The definition (17) can be 
straightforwardly extended to
the functions of $n$ variables  $f{\equiv}f(X_1(z),...X_n(z))$.
The direct although lengthy computation of the commutators
(using the picture-equivalence relations (12)) shows that
seven $T_{-4,n}$-generators of the ghost number $-4$ cohomology (16)
taken with 8 generators of $SU(3)$ of (13) combine into
15 generators of $SU(4)$. As in the case of $SU(3)$ the Cartan subalgebra
 of $SU(4)$ is generated by the zero momentum currents
$L,H$ and $T_{-4,0}$ of (13) and (16).
As previously, this $SU(4)$ algebra intertwines with the conformal symmetry
of the theory. The repeated applications of the 
lowering subalgebra of $SU(4)$
 to the dressed tachyonic vertex lead to the extended
set of the ghost-dependent discrete states which are the multiplets
of $SU(4)$. It is natural to assume that this construction
can be further generalized to include the generators of higher ghost numbers.
The total number of the BRST-invariant generators  $T_{-n,m}$
with the ghost numbers $-N\leq{-n}\leq{-3}$ and the momenta
$-n+1\leq{m}\leq{n-1}$ for each n, combined with three standard
 $SU(2)$ generators, is equal to $N^2-1$. It is natural to conjecture
that altogether they  generate  $SU(N)$, although in
this paper we leave this fact without a proof.
As before, the Cartan subalgebra of SU(N) is generated by
the commuting zero momentum generators 
 $T_{-n,0};n=0;3,4,...,N$.
Applying repeatedly the lowering $SU(N)$ subalgebra ( i.e.
the $T_{-n,m}$'s with $m\leq{0}$) 
 to the dressed tachyon operator
would then generate the extended set of the physical ghost-dependent
discrete states - the multiplets of SU(N).
In the next section we will address the question of BRST-nontriviality
of the $T_{-n,m}$-generators. Finally, we will demonstrate the 
explicit construction
 of new ghost-dependent discrete states from
by $T_{-n,m}$ for the case of $SU(3)$
and compute their structure constants.

\centerline{\bf BRST nontriviality of the $T_{-n,m}$-currents}

The BRST charge of the one-dimensional 
NSR superstring theory is given by the usual worldsheet integral
\eqn\lowen{Q_{brst}=\oint{{dz}\over{2i\pi}}
{\lbrace}cT-bc\partial{c}+\gamma{G_{matter}}
-{1\over4}b\gamma^2\rbrace}
where $G_{matter}$ is the full matter ($c=1+$Liouville) supercurrent.
The BRST-invariance of the $T_{-n,m}$-currents is easy to verify  by 
simple calculation of their commutators with $Q_{brst}$.
Indeed, as the primaries of dimension one and $b-c$ ghost number zero
they commute
with  the stress-tensor part of $Q_{brst}$ and their operator products with
the supercurrent terms of $Q_{brst}$ are all non-singular.
The BRST-invariance of the picture-equivalent currents
$T_{n-2,m}=Z(:\Gamma^{2n-2}cS_{-n,m}:)$,
where $S_{-n,m}$ are the integrands of $T_{-n,m}$,
 simply follows from the invariance of $\Gamma$ and $Z$.

The BRST-nontriviality of these operators 
is less transparent and  needs separate proof.
In principle, the BRST non-triviality
of any of the operators can be proven if one shows that
they produce non-vanishing correlators, which is what will be done
precisely in the next section for the case of $n=3$.
In this section, however, we will present the proof
for general values of $n$, without computing the correlators.
For each $n$, it is sufficient to prove the non-triviality of
$T_{-n,n-1}$-operators, as the currents with the lower momenta are 
 obtained from $T_{-n,n-1}$ by repeated applications of the
lowering generator of $SU(2)$.
The BRST non-triviality of new discrete states the multiplets
of $SU(n)$ then automatically follows from the non-triviality of the 
$T$-currents.
 In other words, we need to show that for each $n$ there are no
 operators $W_n$ in the small Hilbert space such that 
${\lbrack}Q_{brst},W_n\rbrack=T_{-n,n-1}$.
As it is clear from the form of the BRST charge (18) there are only
two possible expressions for $W_n$ 
(up to total derivatives that drop out after the worldsheet integration)
 which commutator with the BRST
charge may produce the $T$-currents:
\eqn\grav{\eqalign{W_n=W_{n}^{(1)}+W_{n}^{(2)}\cr
W_n^{(1)}=\sum_{k=1}^{n-1}\alpha_k
\oint{{dz}\over{2i\pi}}e^{-(n+1)\phi+i(n-1)X}\partial^{(k)}\xi
\partial^{(n-k)}X\cr
W_n^{(2)}=\sum_{k,l=1,k\neq{l}}^{k,l=n;k+l\leq{2n}}\alpha_{kl}
\oint{{dz}\over{2i\pi}}e^{-(n+2)\phi+i(n-1)X}\psi
\partial^{(k)}\xi\partial^{(l)}\xi\partial^{(2n-k-l)}{c}}}
with $\alpha_k$ and $\alpha_{kl}$ being some coefficients
and $\xi=e^\chi$.
Generically, the $W$-operators may also contain the worldsheet derivatives
of the exponents of $\phi$ 
(corresponding to the derivatives of delta-functions of
the ghost fields), but one can always bring these
operators to the form (19) by partial integration.
Clearly, the operators $W_n^{(1)}$ and $W_n^{(2)}$ are the conformal 
dimension one operators satisfying the relations
\eqn\grav{\eqalign{
\lbrack\oint{{dz}\over{2i\pi}}\gamma^2{\psi\partial{X}},W^{(1)}\rbrack\sim
T_{-n,n+1}\cr
\lbrack\oint{{dz}\over{2i\pi}}\gamma^2b,W_n^{(2)}\rbrack\sim{T_{-n,n-1}}\cr
\lbrack\oint{{dz}\over{2i\pi}}\gamma^2{\psi\partial{X}},W^{(2)}\rbrack
=\lbrack\oint{{dz}\over{2i\pi}}\gamma^2b,W_n^{(1)}\rbrack=0}}
Therefore the $T$-currents are BRST-trivial if and only if
there exists at least one combination  of the coefficients $\alpha_k$
or $\alpha_{kl}$ such that $W_n$ commutes with 
the stress tensor part of $Q_{brst}$, that is,
either 
\eqn\lowen{\lbrack\oint{{dz}\over{2i\pi}}(cT-bc\partial{c}),
{W_n^{(1)}}\rbrack=0} or 
\eqn\lowen{\lbrack{\oint{{dz}\over{2i\pi}}(
cT-bc\partial{c}),W_n^{(2)}}
\rbrack=0}
 So our goal is to show that no such combinations exist.
We start with $W_n^{(1)}$. Simple computation of the
commutator, combined with the partial integration 
(to get rid of terms with the derivatives of $\phi$)
gives
\eqn\grav{\eqalign{\lbrack\oint{{dz}\over{2i\pi}},W_n^{(1)}\rbrack=
e^{-(n+1)\phi+i(n-1)X}
\sum_{k=1}^{n-1}\sum_{a=1}^k{{k!}\over{a!(k-a)!}}
\partial^{(a)}c{\lbrace}\partial^{(k-a+1)}\xi\partial^{(n-k)}X\alpha_k
\cr
+\partial^{(n-k)}\xi\partial^{(k-a+1)}X\alpha_{n-k}\rbrace
-{\alpha_k}{\lbrace}n\partial{c}\partial^{(k)}\xi\partial^{(n-k)}X
+c\partial(\partial^{(k)}\xi\partial^{(n-k)}X)}}
The operator $W_n^{(1)}$ is BRST-trivial if for any combination
of the $\alpha_k$ coefficients this commutator vanishes.
The condition $\lbrack{Q_{brst}},W_n^{(1)}\rbrack=0$ 
gives the system of linear constraints
on $\alpha_k$. That is,it's easy to see that the 
right hand side of (23) consists of
the terms of the form $${\sim}e^{-(n+1)\phi+i(n-1)X}
\partial^{(a)}{c}\partial^{(b)}\xi\partial^{(n-a-b+1)}X$$ with
$a,b\geq{1},a+b\leq{n}$. The $T_{-n,n-1}$-operator is BRST-trivial if and only 
if the coefficients (each of them given by some linear combination
of $\alpha_k$) in front of the terms in the
right hand side of (23) vanish for each independent combination of $a$ and $b$.
The number of independent combinations of $a$ and $b$ 
is given by ${1\over2}n(n+1)$ and is equal to the number of constraints
on $\alpha_k$. The number of $\alpha_k$'s is obviously equal to
$n-1$ that is, for $n\leq{-3}$
the number of the constraints is bigger than the number of
 of $\alpha$'s.This means that the system of equations on $\alpha_k$
has no solutions but the trivial one $\alpha_k=0;k=1,...n-1$
and no operators of the $W_{n}^{(1)}$-type satisfying (21)
 exist. Therefore there is no threat of the BRST-triviality of 
the $T_{-n,n+1}$-operator
from this side.
Next, let's consider the case of $W_{n}^{(2)}$ and show that
there is no combination of the coefficients $\alpha_{kl}$
for which (22) is satisfied. The proof is similar
to the case of $W_{n}^{(1)}$.
The number of independent coefficients $\alpha_{kl}$
(such that $k,l\geq{1};k<l$ and $k+l\leq{2n}$) is equal to $N_1=n^2-n$.
The commutator $\lbrack\oint{{dz}\over{2i\pi}}(cT-bc\partial{c}),
W_n^{(2)}\rbrack$
leads to the terms of the form
$$\sim{\oint{{dz}\over{2i\pi}}}e^{-(n+2)\phi+i(n-1)X}\psi
\partial^{(k)}\xi\partial^{(l)}\xi\partial^{(m)}c\partial^{(2n+1-k-l-m)}c$$
with all integer $k,l,m$ satisfying
$k,l\geq{1},m\geq{0};k<l,m<2n+1-k-l-m;k+l+m\leq{2n+1}$.
The total number of these independent terms, multiplied by various
linear combinations of $\alpha_{kl}$, is given by
the sum $N_2=2{\sum_{p=1}^{n-1}}p(n-p)$ and is equal to the number
of the linear constraints on $\alpha_{kl}$. Since for
$n\leq{-3}$ one always has $N_2>N_1$, there are no non-zero
solutions for $\alpha_{kl}$,
 and therefore no operators
of the $W_n^{(2)}$-type exist.
This concludes the proof of the BRST-nontriviality of
the T-currents. Our proof implied that, generically,  the constraints
on $\alpha$'s are linearly independent
In fact, such an independence can be demonstrated straightforwardly
 by the numerical analysis of the systems of linear equations implied by
$\lbrack\oint{{dz}\over{2i\pi}}(cT-b{c\partial{c}}),W_n^{(1),(2)}\rbrack=0$.

\centerline{\bf SU(n) multiplets and their structure constants}
In this section, we will demonstrate the straightforward
construction  of the $SU(N)$ multiplets of the ghost-dependent
discrete states and compute their structure constants in the
 case of $N=3$.
The discrete states  generated by the lowering operators of
the current algebra (14),(15)
 realise various (generically, reducible)
 representations of the SU(3).
 We start with the  decomposition 
of the current algebra (13),(14),(15)
\eqn\lowen{SU(3)=N_{+}\oplus{N_0}\oplus{N_{-}}}
with the operators $L$ and $H$ being in the Cartan subalgebra
$N_0$, the subalgebra $N_{+}$ consisting of 3 operators
$G_{\pm}$ and $G_3$ with the unit positive momentum
and with 3 lowering operators $F_{\pm}$ and $F_3$ with the unit negative
momentum being in $N_{-}$.
This corresponds to the Gauss decomposition
of $SL(3,C)$ which compact real form is isomorphic to SU(3).
Then the full set of the $SU(3)$ multiplet states can be obtained 
simply by the various combinations of the $N_{-}$-operators acting
on the set of the highest weight vector states.
So our goal now is to specify the highest weight vectors.
In fact, it's easy to check that,
just as in the SU(2) case, the highest weight vectors
are simply 
the dressed tachyonic operators 
\eqn\lowen{\oint{{dz}\over{2i\pi}}
V_l(z)=\oint{{dz}\over{2i\pi}}e^{ilX+(l-1)\varphi}(l\psi(z)-i(l-1)\lambda)}
where $l$ is integer valued, however there is
one important subtlety.
It is clear, for example, that the tachyon with $l=1$ cannot be the
highest weight vector since $\lbrack{H,{\oint{{dz}\over{2i\pi}}}V_1}
\rbrack\equiv
\lbrack{H,T_{0,1}\rbrack={1\over{\sqrt{2}}}T_{-3,1}}$, 
therefore
$\oint{{dz}\over{2i\pi}}V_1$ is not 
an eigenvector of $N_0$. In addition, $\oint{{dz}\over{2i\pi}}V_1$ isn't
annihilated ny $N_{+}$ since $\lbrack{T_{-3,1},
\oint{{dz}\over{2i\pi}}V_1\rbrack}\sim{T_{-3,2}}$.
Therefore we have to examine carefully
how the operators of $N_0$ and $N_{+}$ act on generic 
$\oint{{dz}\over{2i\pi}}V_l$.
First of all, it is clear that all the $V_l$'s with $l\geq{2}$
are annihilated by $N_{+}$ since their OPE's with the integrands
of $G_{\pm}$ and $G_3$ are non-singular.
Also, all these tachyons are the eigenvectors of $L$ of $N_0$ with 
the weight $l\over2$ i.e. one half of the isospin value
(the actual isospin value must be taken  equal to $l$
since our conventions involve the factor of $-{1\over2}$ in the stress-energy
tensor and hence our normalization of the $<XX>$-propagator is
one half of the one used in ~{\pol}; 
thus the $L$-operator corresponds to one half of the appropriate
Gell-Mann matrix).
 So we only need to check how these operators are acted on by 
the hypercharge generator $H$ of $N_0$. Simple calculation gives
\eqn\grav{\eqalign{\lbrack{H,{\oint{{dz}\over{2i\pi}}}V_l}\rbrack
=\lbrack
{i\over{3{\sqrt{2}}}}{\oint{{dz}\over{2i\pi}}}e^{-3\phi}
(\partial^2{X}\psi-2\partial{X}\partial\psi),\oint{{dw}\over{2i\pi}}
e^{ilX}(l\psi-i(l-1)\lambda)(w)\rbrack
\cr={i\over{3{\sqrt{2}}}}\oint{{dw}\over{2i\pi}}e^{-3\phi+ilX+(l-1)\varphi}
{\lbrace}3il^2\partial\psi\psi-3l\partial^2{X}+{1\over2}il^2{P_{-3\phi}^{(2)}}
\cr
+(6l\partial{X}-3l(l-1)\psi\lambda)\partial\phi+3l(l-1)\partial\psi\lambda
\rbrace}}
It is convenient to integrate by parts the  terms 
containing the derivatives of $\phi$. Performing
 the partial integration we get
\eqn\grav{\eqalign{\lbrack{H,{\oint{{dz}\over{2i\pi}}}V_l}\rbrack
={i\over{3{\sqrt{2}}}}
\oint{{dz}\over{2i\pi}}e^{-3\phi+ilX+(l-1)\varphi}{\lbrace}
3il^2\partial\psi\psi
-l(1+{{l^2}\over2})\partial^2{X}\cr+il^2(l-1)\partial^2\varphi
+l(l-1)(2\partial\psi\lambda-\psi\partial\lambda)
+(il\partial{X}+(l-1)\partial\varphi)(2l\partial{X}-l(l-1)\psi\lambda)
\cr+{{il^2}\over2}(il\partial{X}+(l-1)\partial\varphi)^2
\rbrace}}
The operator on the right-hand side of (27) is just the dressed tachyon
at the picture $-3$, up to the hypercharge related numerical factor.
To compute the value of the hypercharge we have to picture transform
this operator three times to bring it to the original zero picture.
Since we are working with the integrated vertices, the c-ghost term
 of the local picture-changing operator $\sim{c\partial\xi}$ 
doesn't act on the 
integrand (this can also be seen straightforwardly from the
$\lbrack{Q_{brst};\xi...}\rbrack$-representation of the picture-changing),
while the $b$-ghost term annihilates the right-hand side of (27).
Therefore it is sufficient to consider the matter part of the picture-
changing operator 
\eqn\lowen{\Gamma=:\delta(\beta)G_{matter}:=-{i\over{\sqrt{2}}}e^\phi
(\psi\partial{X}+\lambda\partial\varphi+\partial\lambda)}
Again,  the factor of ${1\over{\sqrt{2}}}$ in the matter supercurrent 
$G_{matter}$
appears because our normalization of the $<XX>$-propagator differs
from ~{\pol} by the factor of $1\over2$.
Applying the picture-changing operator to the right-hand side of
(28)  gives
\eqn\lowen{:\Gamma::\lbrack{H,{\oint{{dz}\over{2i\pi}}}V_l}\rbrack:
=(-{i\over{\sqrt{2}}})({i\over{3\sqrt{2}}})\oint{{dz}\over{2i\pi}}
e^{-2\phi+ilX}\psi(z)\times{(2-2l)}}
i.e. the tachyon at the picture $-2$
Finally, applying the  picture-changing to the right-hand side of 
(29) two times more is elementary and we obtain 
\eqn\lowen{:\Gamma:^3:\lbrack{H,
{{\oint{{dz}\over{2i\pi}}}}V_l(z)}\rbrack={{l(l-1)}\over{6}}
\oint{{dz}\over{2i\pi}}V_l(z)}
i.e. we have proven that the tachyons with the integer momenta $l\geq{2}$
are the highest weight vectors of $SU(3)$. The coefficient
in front of $V_l$ gives the value of the tachyon's SU(3) hypercharge 
\eqn\lowen{s(l)={{l(l-1)}\over{6}}}
Since either $l$ or $l-1$ is even,this guarantees that possible
values of the hypercharge are always the multiples of ${1\over3}$.
Having determined the highest weight vectors we can now easily obtain
the spectrum of the physical states - the multiplets of SU(3).
The vertex operators are simply given by
\eqn\lowen{\oint{{dz}\over{2i\pi}}V_{l;p_1p_2p_3}=
F_{+}^{p_1}F_{-}^{p_2}F_3^{p_3}\oint{{dz}\over{2i\pi}}V_l(z)}
with all possible integer values of $p_1,p_2$ and $p_3$
such that $p_1+p_2+2p_3\leq{2l}$
This construction is in fact isomorphic
to the Gelfand-Zetlin basis of the irreps of $SU(3)$, or
 to the tensor representations
of $SU(3)$ on the $T^P$-spaces generated by the polynomials of
degree $P=p_1+p_2+p_3$ in three variables $x_1,x_2,x_3$ spanned by 
$x_1^{p_1}x_2^{p_2}x_3^{p_3}$, with $x_{1,2,3}$ being the covariant
vectors under $SU(3)$ ~{\prakash}. In particular,
the values of $P$ can be used to label the irreducible representations.
In  our case, the values of $p_1$, $p_2$, $p_3$ and $P$
 can be easily related
to the isospin projection $m$, the hypercharge $s$ and
the ghost cohomology numbers $N$ of the vertex operators (32).
Applying $L$ and  $H$ to the operators (32)
and using the commutation relations  (14) we get
\eqn\grav{\eqalign{
s=p_1-p_2+{{l(l-1)}\over6}\cr
m=l-p_1-p_2-2p_3\cr
N_{max}=-3P=-3(p_1+p_2+p_3)}}
where $N^{max}=-3P$ is the biggest (in terms of the absolute value)
ghost cohomology number of the operators appearing in the expression for
$V_{l;p_1p_2p_3}$.

Conversely,
\eqn\grav{\eqalign{p_1=-{1\over2}(m-l+s)-{1\over3}N^{max}-{1\over{12}}
l(l-1)\cr
p_2={1\over2}(m-l-s)-{1\over3}N^{max}+{1\over{12}}l(l-1)\cr
p_3=l-m+{1\over3}N^{max}}}
The SU(3) symmetry then makes it straightforward to determine the 
structure constants of the operators (32), given by the three-point 
correlators 
\eqn\lowen{A(l_{1,2,3};p_{1,2,3};q_{1,2,3};r_{1,2,3})=
<cV_{l_1;p_1p_2p_3}cV_{l_2;q_1q_2q_3}cV_{l_3;r_1r_2r_3}>}
The symmetry determines these correlators
up to the overall isospin function $f(l_1,l_2)$ which we will fix later.
As for the $m$ and $s$-dependences of the correlator, they are completely
fixed
by the $SU(3)$ symmetry and, accordingly, by the $SU(3)$ Clebsch-Gordan 
coefficients.
It is now straightforward to deduce the structure constants
by using the relations (33),(34) and the SU(3) Clebsch-Gordan
decomposition:

\eqn\grav{\eqalign{
|P;lms>=\sum_{{\lbrace}Q,R;|Q-R|\leq{P}{\rbrace}}
\sum_{{\lbrace}l_1,l_2;|l_1-l_2|\leq{l}
{\rbrace}}
\sum_{\lbrace\mu,\sigma\rbrace}\cr
{D^{Q;R|P}_{l_1\mu\sigma;l_2,m-\mu,s-\sigma|lms}}
{\times}|Q;l_1\mu\sigma>|R;l_2(m-\mu)(s-\sigma)>\cr
P=p_1+p_2+p_3\cr
Q=q_1+q_2+q_3\cr
R=r_1+r_2+r_3}}

where  $|...>$ are the eigenfunctions of the isospin, its
projection and of the hypercharge,
${D^{Q;R|P}_{l_1,\mu,\sigma;l_2,m-\mu,s-\sigma|lms}}$
are the $SU(3)$ Clebsch-Gordan coefficients.
Furthermore, it is convenient to write the SU(3) Clebsch-Gordan
coefficients in the form:
\eqn\lowen{{D^{Q;R|P}_{l_1,\mu,\sigma;l_2,m-\mu,s-\sigma}}
=\alpha^{l_3,s;R,P,Q}_{\sigma;{l_1},l_2}{\times}C^{l_1,l_2,l_3}_{\mu,m-\mu,m}}
where $C^{l_1l_2l_3}_{\mu,m-\mu,m}$ are the usual SU(2) Clebsch-Gordan
coefficients and $\alpha^{l_3sRPQ}_{\sigma{l_1}l_2}$ are the $U(1)$
isoscalar factors computed in ~{\grig}.
The three-point amplitude (35) is then given by
\eqn\grav{\eqalign{A(l_{1,2,3};p_{1,2,3};q_{1,2,3};r_{1,2,3})
=f(l_1,l_2)D^{P,Q|R}_{{{l_1}},m_1,s_1;{l_2}
m_2s_2|{{l_3}},m_1+m_1,s_1+s_2}
=f(l_1,l_2)\cr\times{\alpha^{l_3,(p_1-p_2+q_1-q_2+{1\over6}(l_1(l_1-1)+l_1
(l_2-1)));{R},{P},{Q}}_{p_1-p_2+
{{l_1(l_1-1)}\over6};{{l_1}\over2},{{l_2}}}}
{C^{{{l_1}},{{l_2}},
{{l_3}}}_{{{l_1-p_1-p_2-2p_3}},{{l_2-q_1-q_2-2q_3}},
{{l_3-r_1-r_2-2r_3}}}}}}
where 
\eqn\grav{\eqalign{
P=p_1+p_2+p_3=-{1\over3}N_1^{max}\cr
Q={q_1+q_2+q_3}=-{1\over3}N_2^{max}\cr
R=r_1+r_2+r_3=-{1\over3}N_3^{max}\cr
l_1+l_2-p_1-p_2-2p_3-q_1-q_2-2q_3=l_3-r_1-r_2-2r_3 
}}
The final step is to determine the function $f(l_1,l_2)$.
To find it we will follow the procedure quite similar to ~{\pol}
and consider the special case of $p$,$q$ and $r$ when the 
computation of the OPE can be done explicitly; then we will compare
the result
with the general formula (38) in order to fix $f(l_1,l_2)$.
Namely, consider the OPE 
 \eqn\lowen{V_{l_1;010}(z)V_{l_2;000}(w)=
:F_{-}e^{il_1X+(l_1-1)\varphi}:(z):e^{il_2X+(l_2-1)\varphi}:(w)}
To evaluate the structure constants of this OPE we note that
the all the operators appearing on the right-hand side
carry the momentum $p_x=l_1+l_2-1$ in the $X$-direction and
$p_L=l_2+l_1-2$ in the Liouville direction.
But the only primary  field of dimension $1$
with such a property is (up to the picture-changing) the
 highest weight vector (25) with $l=l_1+l_2-1$.
Therefore we have
\eqn\lowen{V_{l_1;010}(z)V_{l_2;000}(0)\sim{A\over{z}}V_{l_1+l_2-1;000}(0)}
and we only need to determine the structure constant $A$ in order to fix
$f(l_1,l_2)$.
On the left hand side, it is convenient to take the highest weight vectors at 
the ghost picture $-1$:
$$V_{l;000}\equiv{V_l}=-{\sqrt{2}}
e^{-\phi+ilX+(l-1)\varphi}; l=l_1,l_2$$
where the normalization factor of $-{\sqrt{2}}$ ensures 
that applying the picture-changing operator $\Gamma$ to 
$V_{l;000}$ at picture $-1$ gives  $V_{l;000}$ at picture zero 
 normalized according to (25).
As $F_{-}={1\over2}T_{0,-1}-{1\over{2{\sqrt{2}}}}T_{-3,-1}$,
we start with the evaluation of the OPE of the  
$:T_{0,-1}e^{-\phi+il_1X+(l_1-1)\varphi}:$ part of $V_{l_1;100}$
with $V_{l_2;000}$.
We have:
\eqn\grav{\eqalign{:T_{0,-1}V_{l_1;000}:(z)V_{l_2;000}(0)\cr=
2\oint{{du}\over{2i\pi}}:e^{-iX}\psi:(u+z)
:e^{-\phi+i{l_1X}+{(l_1-1)}\varphi}:(z)
e^{-\phi+il_2X+(l_2-1)\varphi}(0)
\cr=e^{-2\phi+i(l_1+l_2-1)X+(l_1+l_2-2)\varphi}\psi(0)
\oint{{du}\over{2i\pi}}u^{-l_1}(u+z)^{-l_2}z^{l_1+l_2-2}}}

Introducing the integration variable $w={u\over{z}}$ have:
\eqn\grav{\eqalign{2\oint{{du}\over{2i\pi}}u^{-l_1}(u+z)^{-l_2}
z^{l_1+l_2-2}={2\over{z}}\oint{{dw}\over{2i\pi}}
w^{-l_1}(1+w)^{-l_2}={2\over{z}}\times{{(l_1+l_2-2)!}\over{(l_1-1)!(l_2-1)!}}}}
Transforming the right-hand side to the zero picture by the operator (28)
and noting that
$$:\Gamma^2:\oint{{dz}\over{2i\pi}}e^{-2\phi+i(l_1+l_2-1)X+(l_1+l_2-2)\varphi}
\psi(z)=-{{l_1+l_2-1}\over2}\oint{{dz}\over{2i\pi}}V_{l_1+l_2-1;000}(z)$$
 we get:
\eqn\grav{\eqalign{:T_{0,-1}V_{l_1;000}:(z)
:V_{l_2;000}:(0)\sim
-{1\over{z}}{{(l_1+l_2-1)!}\over{(l_1-1)!(l_2-1)!}}V_{l_1+l_2-1;000}(0)}}
The next step is to compute the OPE between
$T_{-3;-1}V_{l_1;000}$ and $V_{l_2;000}$.

To simplify the calculation, we shall take the generator
$T_{-3;-1}$ at the picture $+1$ by using the equivalence relations
(12) and the  operators $V_{l_1;000}$  and $V_{l_2;000}$ both at the picture 
$-1$ : $V_{l_{1,2};000}=-{\sqrt{2}}
e^{-\phi+il_{1,2}X+(l_{1,2}-1)\varphi}$.
Note that, as the expressions the operators $V_{l_{1,2};000}$ at the picture 
$-1$ do not  contain any $b$ and $c$ ghost fields,
we can as well disregard the $b-c$ ghost part
of $T_{+1,1}$ (since the form of the OPE (41) is fixed, the interaction of
this part with the highest weight vectors can only contribute
BRST-trivial terms)
  
We have:
\eqn\grav{\eqalign{:T_{+1;1}V_{l_1;000}:(z)V_{l_2;000}(0)
\cr=2\oint{{du}\over{2i\pi}}e^{\phi-iX}{\lbrace}\partial{\psi}\psi
+{1\over2}(\partial{X})^2-{i\over2}\partial^2{X}\rbrace(u)
e^{-\phi+il_1{X}+(l_1-1)\varphi}(z)e^{-\phi+il_2X+(l_2-1)\varphi}(0)
\cr=2e^{-\phi+i(l_1+l_2-1)X+(l_1+l_2-2)\varphi}(0)
\oint{{du}\over{2i\pi}}
u^{-l_1+1}(u+z)^{-l_2+1}z^{l_1+l_2-2}\cr\times
\lbrace{{l_1-l_1^2}\over{2u^2}}+{{l_2-l_2^2}\over{2(u+z)^2}}
-{{l_1l_2}\over{u(z+u)}}\rbrace
\cr=
-{\sqrt{2}}e^{i(l_1+l_2-1)X+(l_1+l_2-2)\varphi}((l_1+l_2-1)\psi
-i(l_1+l_2-2)\lambda)(0)
\cr\times\oint{{du}\over{2i\pi}}
u^{-l_1+1}(u+z)^{-l_2+1}z^{l_1+l_2-2}
\lbrace{{l_1-l_1^2}\over{2u^2}}+{{l_2-l_2^2}\over{2(u+z)^2}}
-{{l_1l_2}\over{u(z+u)}}\rbrace}}
where the terms inside the figure brackets
in the integral are due to the contractions
of the derivatives of $X$ in the expression for
$T_{1,-1}$ with the exponents in $V_{l_{1,2};000}$
(note that the term with $\partial\psi\psi$ 
of $T_{1,-1}$ cannot contribute
anything but the BRST-trivial part because of
 the fixed form of the OPE (41))
Performing the contour integration precisely is in (42)-(44)
we obtain
\eqn\grav{\eqalign{:T_{1,1}V_{l_1;000}:(z)
V_{l_2;000}(0)=-{{\sqrt{2}}\over{z}}V_{l_1+l_2-1;000}
\cr\times\lbrace{{(l_1-l_1^2)}\over{2}}{{(l_1+l_2-1)!}\over{l_1!(l_2-2)!}}
+{{(l_2-l_2^2)}\over{2}}{{(l_1+l_2-1)!}\over{l_2!(l_1-2)!}}
-l_1l_2{{(l_1+l_2-2)!}\over{(l_1-1)!(l_2-1)!\rbrace}}
\cr=
-{{\sqrt{2}}\over{z}}V_{l_1+l_2-1;000}
\times{{(l_1+l_2-2)!}\over{(l_1-1)!(l_2-1)!}}
(l_1+l_2-1-l_1l_2)
}}
Finally, collecting together (44) and (46) 
we get
\eqn\grav{\eqalign{F_{-}V_{l_1;000}(z)V_{l_2;000}(0)
=-{{l_1l_2}\over{2z}}{{(l_1+l_2-2)!}\over{(l_1-1)!(l_2-1)!}}
V_{l_1+l_2-1;000}}}
Comparing the operator product (47) with the general
expression (38) for $q_{1,2,3}=r_{1,2,3}=0$,
$p_1=p_3=0$ and $p_2=1$ we can now  easily fix the isospin
function $f(l_1,l_2)$ to be equal to
\eqn\grav{\eqalign{f(l_1,l_2)=
-{{l_1l_2}\over{2}}{{(l_1+l_2-2)!}\over{(l_1-1)!(l_2-1)!}}
(\alpha^{l_1+l_2-1,
{{l_1(l_1-1)+l_2(l_2-1)}\over{6}};0,1,0}_{{{l_1(l_1-1)}\over{6}}-1;l_1,l_2}
\times{C^{l_1,l_2,l_1+l_2-1}_{l_1-1,l_2,l_1+l_2-1}})^{-1}\cr
={1\over2}l_1l_2{\sqrt{{{l_1+l_2}\over{l_2}}}}
{{(l_1+l_2-2)!}\over{(l_1-1)!(l_2-1)!}}
(\alpha^{l_1+l_2-1,
{{l_1(l_1-1)+l_2(l_2-1)}\over{6}};
0,1,0}_{{{l_1(l_1-1)}\over{6}}-1;l_1,l_2})^{-1}
}}
This concludes the computation of the structure
constants for the ghost-dependent discrete states of
SU(3) multiplet.

\centerline{\bf Conclusion and discussion}
In this paper we have shown that the interaction
of the 2-dimensional supergravity with the $\beta-\gamma$ system of the 
superconformal ghost crucially extends the spectrum of the physical
states and enhances the underlying symmetry of the theory.
We have shown  that the discrete states generated by the currents
of the ghost number cohomologies
  such that $-N\leq{n}\leq{N-2};N\geq{3}$,combined with the
standard SU(2) discrete states, form the $SU(N)$ multiplets.
We have demonstrated this explicitly for $N=3$ and $4$ and 
conjectured for higher
values of $N$.

In the $N=3$ case structure constants of the ghost-dependent discrete states
of the SU(3) multiplet are determined by the appropriate Klebsch-Gordan
 coefficients. Since $SU(3)$ is isomorphic to ${SL(3,R)}$, 
the operator algebra
 (38), (48) is related to the algebra of 
the volume preserving diffeomorphisms
in $3$ dimensions after the appropriate rescaling of the vertices,
 just as the OPEs of the standard $SU(2)$ discrete states
lead the algebra of the area preserving diffeomorphisms on the plane
after the redefinition of the fields
~{\pol}. 
Physically, the volume preserving diffeomorphisms 
$sDiff(3)$ in three dimensions describe the dynamics of
ideal incompressible fluid in $d=3$.
In this context, the appearance of the CG coefficients in the
structure constants is quite natural, as the vertex operators
(32) can be interpreted as Clebsch variables for the space of vorticities
~{\arnold, \mars}. The beta-functions
of the $SU(3)$ multiplet states would then generate
the RG flows similar to the dynamics of the co-adjoint orbits considered in 
~{\mars}.
From this point of view, the RG equations are  the Lie-Poisson equations
associated to $sDiff(3)$.
Unfortunately, because of the complexity of the expressions for 
the SU(3) CG coefficients, as well as due to difficulties
 in the classification of the volume preserving diffeomorphisms in 
higher dimensions, the precise form of the necessary field rescaling
appears to be far more complicated then in the $SU(2)$ case.
We hope to elaborate on it in details
in the future work.

Generalizing these arguments to the ghost cohomologies of higher ghost
numbers,
it is seems natural to conjecture that the structure constants
of the $SU(N)$ multiplets are related to
the Clebsch-Gordan coefficients of SU(N), up
to the functions of the Casimir eigenvalues which should be determined
similarly to the isospin function $f(l_1,l_2)$ of
(48), by considering the special cases
of the operators when straightforward    computation of the OPEs
is accessible. The operator algebra should be related then to
the volume preserving diffeomorphisms in $N$ dimensions and
the vertex operators of the $SU(N)$ multiplet could be 
interpreted in terms of the Clebsch variables
of the vortices in $N$-dimensional incompressible liquid.
The case of the special interest is $N=4$;
the ghost-dependent discrete states are then the $SU(4)$ multiplets,
i.e. are in the representation of the conformal group in $d=4$.
The currents generating the $SU(4)$ algebra could then be related to
$d=4$ conformal generators. It would be interesting to relate them to the
twistor variables of $d=4$ twistor superstrings ~{\witten, \wb}.
Particularly, the physical vertices of $d=2$ closed
 superstrings
with the holomorphic and antiholomorphic 
parts being 
in the ghost cohomologies of negative numbers $-n$  with $|n|{\leq}4$,
combined with the standard $SU(2)\oplus{SU(2)}$ primaries,
are the multiplets of $SU(4)\oplus{SU(4)}$,
isomorphic to the bosonic part of $PSU(2,2|4)$, the full symmetry
group of open twistor superstrings and of the $d=4$ super Yang-Mills theory.
 All this implies that the $c=1$ theory coupled 
to the $\beta-\gamma$ ghosts can be used to describe field theories in
 higher dimensions. The question is where the extra dimensions come from.
This question will be studied in details in our next paper, while
in this work we shall give only the rough qualitative explanation. 
In order to answer this question one has to carefully analyze
the $\beta$-functions of the $SU(N)$ multiplets.
The important subtlety here  is that the structure constants
computed in (38), (48), do not fully determine the
$\beta$-functions, as they
involve only the states from the ghost cohomologies of negative
numbers. However, because of the equivalence relations (12) there are also 
the positive ghost number representations for these states.
That is, the relations (12) 
imply the isomorphism between the ghost cohomologies
of negative and positive numbers: 
\eqn\lowen{G_{-N}\sim{G}_{N-2}}
One consequence of this isomorphism is that the
 OPEs of the ghost-dependent discrete states are,
generally speaking,
picture-dependent.  
The illustrate the picture dependence of the OPEs, consider the 
following simple example.
Consider the OPE of two operators $V_1=\oint{{dz}\over{2i\pi}}W_1(z)$ and 
$V_2=\oint{{dz}\over{2i\pi}}W_2(z)$
in the ghost number cohomologies $-N_1\leq{-3}$ and $-N_2\leq{-3}$.
For simplicity,let us take them in their maximal existing pictures
 that is, $-N_1$ and $-N_2$ and
consider the special case of $N_2-N_1\geq{3})$.
By using the equivalence relations (12)
 similar to the reflection identities of the Liouville theory
~{\dorn, \zam}, let us first
take $V_1$ at  the $positive$ picture $N_1-2$ and
leave $V_2$ at the $-N_2$-picture
(recall that $V_1$ exists in all the pictures below $-N_1$
and all above $N_1-2$; similarly for $V_2$).
The necessary equivalence transformation is given by
\eqn\lowen{V_1\rightarrow{Z}(:\Gamma^{2N_1-2}cW_1:)}
Then the OPE of $V_1$ and $V_2$ consists of the operators
of ghost number cohomologies up to $G_{N_1-N_2-2}\sim{G}_{N_2-N_1}$
(given the equivalence relation (12))
Conversely, let us take $V_2$ at the picture $N_2-2$ by the equivalence
transformation
\eqn\lowen{V_2\rightarrow{Z}(:\Gamma^{2N_2-2}cW_2:)} 
and leave $V_1$ at $-N_1$
The OPE of  $V_1$ and $V_2$ would then consist of operators
from the ghost number cohomologies up to
$G_{N_2-2-N_1}\sim{G}_{N_1-N_2}$
 Thus, unlike the first OPE, the latter operator product skips
 the operators of the ghost number cohomologies
 $G_{N_2-N_1-1}\sim{G_{N_1-N_2-1}}$
and $G_{N_2-N_1}\sim{G_{N_1-N_2-2}}$. These operators are simply
screened off by  the extra powers of $\Gamma$ in the equivalence 
transformation
for $V_2$. In the functional integral (which sums over all the ghost pictures)
such a picture assymmetry would generally lead to the Non-Markovian
stochastic terms,or the explicit dependence 
of the $\beta$-function on the operator-valued
worldsheet
variables, as it has been explained in ~{\self}. 
Typically, such $\beta$-function equations have the
 form of either the Langevin stochastic equations,
or of  equations for  incompressible fluid with the random force ~{\self}.
The role of the noise is played
by the worldsheet integrals of ghost-dependent vertex operators
cut of at a scale $\Lambda$. In analogy with 
 the stochastic quantization approach,
one can interpret the effective stochastic time $\Lambda$ as the 
extra dimension of the theory.
Roughly speaking, the equations of the  RG flows
induced by the ghost-dependent states
of $SU(N)$ multiplet $(N\geq{3})$, 
describe the non-Markovian process with
$N-2$ stochastic times, effectively
bringing the $N-2$ extra dimensions to the theory.
In other words, each new ghost cohomology of the underlying
current algebra of the $T$-operators brings an extra
stochastic time in the Non-Markovian RG flow and, subsequently,
an extra dimension.
This qualitatively
 explains the appearance of the extra dimensions but of course
at this point the explanation is still heuristic and the
entire question is yet to be explored in details.

We conclude this paper by noting  that
the operator algebra (38),(48) particularly entails some remarkable
relations concerning the decomposition rules for
 the ghost cohomologies of 
negative numbers.
Consider again the product of $V_1$ and $V_2$ discussed above.
All of the operators on the right hand side of this OPE
have the same ghost number $-N_1-N_2$ but, generally speaking, come
from very different ghost cohomologies.
The relations (38),(48) thus define 
the precise decomposition rules for the product of two
 cohomologies. As we have shown, these decomposition rules are regulated
by $SU(N)$ Clebsch-Gordan coefficients, the fact pointing at
the underlying relations between ghost numbers and spaces of vorticities.
In this context,
the ghost cohomologies may be useful for the classification
of vortices in an incompressible fluid.

\centerline{\bf Acknowledgements}
It is a pleasure to thank A.M. Polyakov, W. Sabra and J. Touma for very useful
discussions. I particularly thank J. Touma for pointing out to me 
the reference ~{\mars}. I also would like to  acknowledge
the hospitality of the Institut des Hautes Etudes Scientifiques
(IHES) in Bures-sur-Yvette where the initial stage of this work was 
completed.

\listrefs
\end